\documentclass[conference]{IEEEtran}    
\usepackage{algorithm}
\usepackage{amsmath}
\usepackage[noend]{algpseudocode}
\usepackage{biblatex}
\usepackage{graphicx}
\usepackage{float}

\begin{document}

\title{Rapid solution for searching similar audio items}
\author{\IEEEauthorblockN{Kastriot Kadriu}
\IEEEauthorblockA{Faculty of Computer and Information Science\\
University of Ljubljana\\
Email: kk5222@student.uni-lj.si
}}
\maketitle
\begin{abstract}
    A naive approach for finding similar audio items would be to compare each entry from the feature vector of the test example with each feature vector of the candidates in a k-nearest neighbors fashion. There are already two problems with this approach: audio signals are represented by high dimensional vectors and the number of candidates can be very large - think thousands. The search process would have a high complexity.
    Our paper will treat this problem through hashing methodologies more specifically the Locality Sensitive Hashing. \\ This project will be in the spirit of classification and clustering problems.
    The computer sound production principles will be used to determine which features that describe an audio signal are the most useful. That will down-sample the size of the feature vectors and speed up the process subsequently. 
\end{abstract}
\section{Introduction}
Audio-visual data is represented by high dimensional vectors. Searching and comparing distances in such large collections has a large complexity. The k-NN problem represents a naive approach for collecting near (similar) items. For low dimensions, k-d trees can be used to answer queries in logarithmic and linear space. In high dimensions, the k-NN problem suffers from the \textit{curse of dimensionality} where either the running time or the space requirements grow exponentially in dimension $d$. \\
Locality Sensitive hashing (LSH) technique consists of projecting each feature vector onto a random distribution of values resulting into random buckets. This random projecting, which is not so random, is repeated several times. The theorem follows the principle that similar items end up in similar buckets. These buckets, that can be represented by hash values, allow us for near-constant time accession of results that represent the near items of a given query example that is also hashed in a similar manner.

\section{Related work}
A locality sensitive hashing model has been initally proposed in \cite{1} where we have a direct approach for reducing computation complexity caused by a high dimensional data. Multiple versions of LSH have since been developed. ANN in \cite{anna} is an efficient library for querying approximate near neighbors in \textit{abritrarily high dimensions}. It implements different data structures based on kd-trees and box-decomposition trees and it employes a couple of different strategies. MinHash \cite{4} is often seen as an instance of locality sensitive hashing. It is a technique for estimating the similarity of two sets in an efficient way, mainly by using the Jaccard similarity coefficient. It is highly used for detecting duplicates of documents, using the so-called shingles. \\
SimHash \cite{3} is a version of LSH used for similar purposes, for finding similar items. It has lower memory requirements than MinHash, however its similarity detection is limited in large distances. \\
One variant of LSH is based on bands from which we are also able to fingerprint (find the uniqueness) of one item that we can later use in identifying duplicate or similar other items.

\section{Locality-sensitive hashing}
Locality-sensitive hashing is a set of techniques that consist of grouping similar items through various metrics that eventually speed up the querying process for near neighbors, or in some cases near-duplication detection, of data. \\
The LSH model is a geo-spatial structure that contains the points (items) that have been projected onto a random distribution. \\
The projected matrix is equal to: \[
[Projected(P)]_{k \times n} = [Random(R)]_{k \times d} [Original(D)]_{d \times n}
\]

The random matrix is generated by using a Gaussian distribution.\\
After projection, we check whether our query point lies above or below the hyper-planes defined by the random projection points, denoted by 1 and 0 respectively. After that, we deal with a matrix in which we perform some basic bit-wise operations to generate the hash value, although the opportunities here are not strict as long as the proper similarity between item is able to be kept by the generated hashes.\\
After hashing items into buckets, it's possible that two similar items are close to each other but on the opposite sides of the border. To handle this issue, the random projection (and hashing) is repeated for a selected number of times. This ensures that similar items will fall into the same bucket at least once. \\
The number of buckets that contain the points is $\approx 2^{s}$ where $s$ is the hash size.
A LSH model can be configured with other hashing functions that are based on deterministic projections. However, random projections are desirable because they lead us to a data-independent evaluation of the model.
\begin{figure*}
    \centering
    \includegraphics[width=15cm]{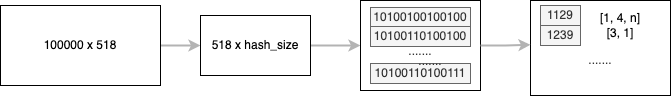}
    \caption{LSH model architecture}
    \label{fig:lsh_arch}
\end{figure*}
\section{Methodology}
\subsection{Dataset}
The dataset used in this project is the Free Music Archive (FMA) dataset \cite{fma}, \cite{fma_paper}. This extensive dataset contains metadata and pre-computed features for 106,574 tracks from 16,341 artists and 14,854 albums arranged in a hierarchical taxonomy of 161 genres. The dataset offers features that describe different properties of the audio. The ones that we will be dealing with are the ones that are computed by Librosa. Librosa decomposes audios into features that describe the following properties such as Chroma Energy Normalized (CENS), Mel-frequency cepstral coefficients, Tonnetz, and Spectral Contracts. For each of these properties, the following values are reported: kurtosis, max, min, mean and standard deviation. 
\subsection{The model}
A working model in the form of a script has been implemented on Python. The model contains three classes: FastAudioSearch, LSH and Table. FastAudioSearch encapsulates the integration and sets up the train and testing procedures. Parameters such as hash size, number of tables and data directory can be passed when running the script.\\
The training process consists of setting up the geo-spatial structure where each item (row) is hashed into a bucket, as shown in Algorithm 1. 
\begin{algorithm}
  \caption{Algorithm for generating LSH given multi dimensional data}
  \begin{algorithmic}[1]
      \Function{build\_lsh\_tables}{$P$}
          \For{table $t$ in tables $T$}
            \State generate random projections
          \EndFor
            \For{table $t$ in tables $T$}
              \State generate hashes using dot projection of $P$ and random projections of $t$
              \State organize document in buckets defined by the respective hashes
            \EndFor
      \EndFunction
  \end{algorithmic}
\end{algorithm}

\begin{algorithm}
  \caption{Algorithm for querying similar items in a LSH model}
  \begin{algorithmic}[2]
      \label{euclidian}
      \Function{query\_lsh\_tables}{$D$, $k$}
            \For{table $t$ in tables $T$}
              \State $H \gets$ generate hashes using dot product of $D$ and random projections of $t$
              \For{test point $d$ in $D$}
                \If{$t$ contains a bucket hashed by $H_{d}$}
                  \State \textbf{return} distances of $i$ with points in $t$ as near neighbor candidates
                \EndIf
            \EndFor
            \EndFor
            \For{test point $d$ in $D$}
                \State $sorted\_dists \gets$ sort distances of d 
                \For{each candidate point $c$ in $sorted\_dists$}
              \State \textbf{return} $c$ only if it has the lowest distance with $d$ in any table
              \EndFor
            \EndFor
      \EndFunction
  \end{algorithmic}
\end{algorithm}

The testing process consists of generating the hashes for the query item and getting the buckets with those hashes from the geo-spatial structure - Algorithm 2. 
Table class contains the hashing procedure for our data whereas LSH sets up the strategy for hashing and retrieving results from multiple tables.\\

\subsection{Complexity}
In the brute force method of the k-NN problem training is considered a constant operation since in this approach there's not an actual training process happening, whereas the querying of the results has a complexity of \textbf{$O(n \times d \times k)$}.
In the k-d tree method, building the tree represents the training phase that has the complexity of \textbf{$O(d \times n \times logn)$}. Usually, we don't care much about the training times since the process isn't expected to be repeated much, but as we can see, in k-d trees the process is dependent on the dimension size and for large dimension, we can see how it can effect the time. The querying times, representing the prediction phase, with this approach are reduced to \textbf{$O(k \times logn)$}. When the dimensionality is big, those trees break down and we end up testing all the nodes, bringing the complexity to \textbf{$O(k \times logn)$}.\\

In the case of LSH with random projections, considering a query, table lookup by its hash value is done in constant time. The process of checking the distances with the near neighbor candidates with the same hash value is $O(d \times \frac{n}{2^{h_s}})$. 
The generation of the hash value is done by taking the dot product of the vector of the query point with the random projection and it has a complexity $O(d \times h_s)$. 
This process is repeated for $t$ number of tables. So the overall complexity for querying approximate near neighbors for a point is considered to be $O(t \times ((d \times h_s) + (d \times \frac{n}{2^{h_s}}))$.

\section{Experiments}
\begin{figure}
\centering
    \includegraphics[width=8cm]{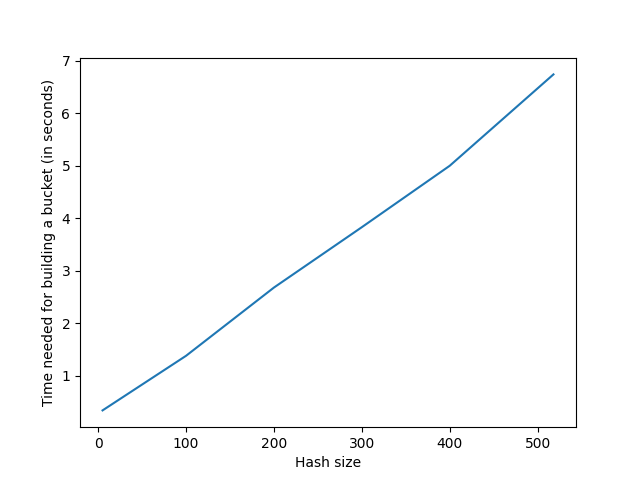}
    \caption{Average time for hashing items in buckets in a single table for different hash sizes, (data size = 100k).}
    \label{fig:experiment1}
\end{figure}

Experiments have been done on a Macbook Pro late 2013 machine running on a quad-core Intel i7 processor and 16gb of RAM.
 The aim of the experiments is to report on several properties of the model. First, we show that hashing the content on blocks for a selected number of tables is a relatively rapid process. Next we show that the random projection is a safe process in which most important data attributes can be saved. We also show that querying similar items candidates for a test example is a robust process and finally we show that those candidates are indeed similar with the queried example. \\ 
 For testing the speed of buckets building, we time our operations. \\
 Since the dataset doesn't provide a collection of ground truth near items that can be used for evaluating the accuracy of the model, we don't use a traditional train-test split. Instead, a sequence of random samples that de-facto represent our testing data is picked from the original dataset. This sequence remains the same for all experiments reported in this paper. 
 \\\\
 The average time for building buckets for a table with different hash sizes is shown in Fig. \ref{fig:experiment1}. The process for building the table includes the dot product with the randomly generated matrix, hashing each row (item) and organising buckets in a dictionary data structure. Subsequently, the training time is increased linearly proportional to the number of tables. \\ \\
 The random projection matrix is controlled by the hash size value that determines the downsampled dimensions. However, if we downsample the original matrix too much, we risk losing important data. For a randomly selected sample, we test how accurate the hash size parameter selection is by testing if the predicted similar items candidates do indeed hold the distance relation with each other when we check their actual distance in the original data matrix in the Euclidean space. This means, that we are looking for a monotonically increasing sequence of distances between the query item and candidates. This process is repeated for 100 re-iterations with a different random sample each time. Then the values are averaged. \\

\begin{figure}
\centering
    \includegraphics[width=8cm]{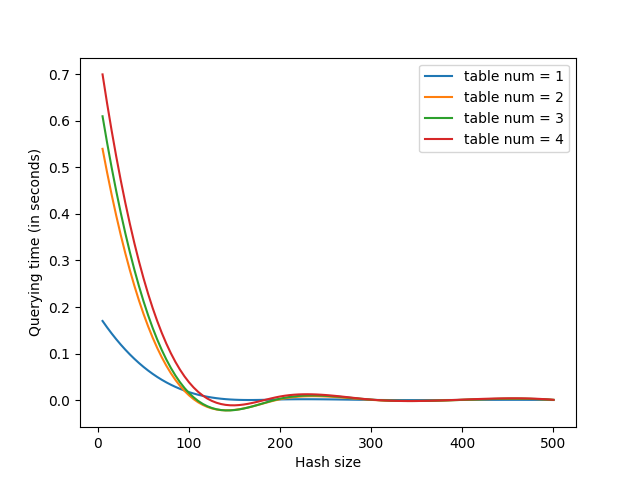}
    \caption{Query processing time for different number of tables, (data size = $100k$, averaged per $100$ queries).}
    \label{fig:experiment2}
\end{figure}
 \begin{figure}
\centering
    \includegraphics[width=8cm]{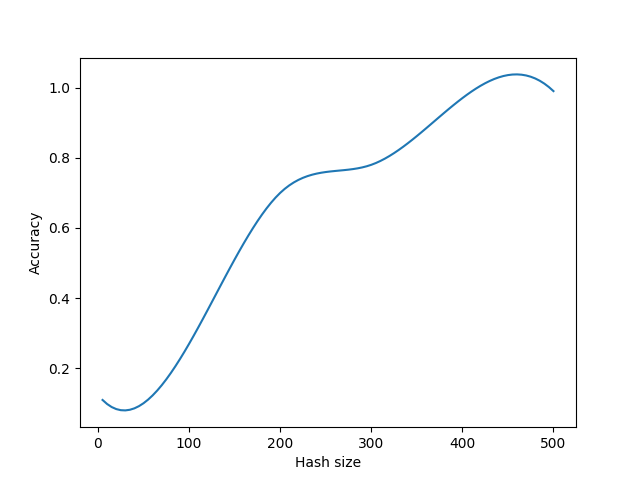}
    \caption{Accuracy of the model evaluated for different hash size values.}
    \label{fig:experiment3}
\end{figure}
\begin{figure}
\centering
    \includegraphics[width=8cm]{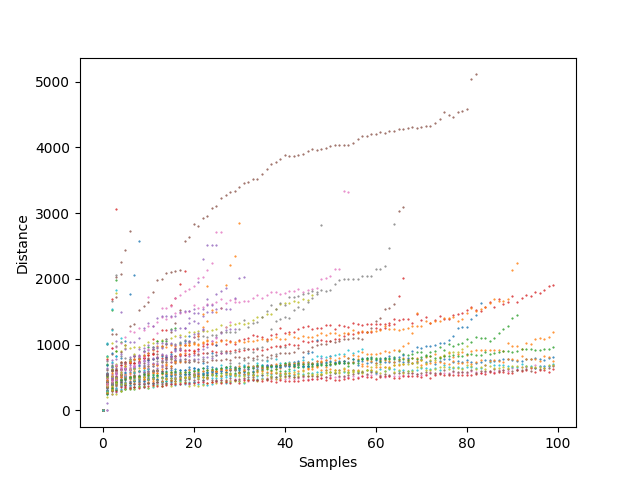}
    \caption{Distance relation of our randomly selected samples, $h^{s}=250, t=1$.}
    \label{fig:experiment4}
\end{figure}

 A sensible hash size parameter is important to be selected because we want to optimize accuracy (higher value) and speed (lower value). \\\\
 \section{Discussions}
From the experiments seen in the previous section, we can observe several properties. \\
First, a lower hash size leads to faster training times (Fig. \ref{fig:experiment1}). That is because of the small number of buckets that are generated and subsequent less operations needed to process the query. But on the other hand, a lower hash size leads to longer querying times because a small number of buckets means a large list of items in those buckets that is retrieved and needs to be sorted. In addition, the data projected in very small dimensions loses the distance relation between items thus affecting the accuracy in retrieving ordered similar items (Fig. \ref{fig:experiment3}).  \\ 
A higher hash size guarantees higher accuracy but with a cost in time. Fine tuning the hash size value altogether with a sensible number of tables can achieve good results for our purposes. \\
From Fig. \ref{fig:experiment1}, we can see also that the time for building tables is increased linearly with the hash size but that should not hinder the process much because the model is supposed to be built (trained) only once.\\ 
In Fig. \ref{fig:experiment2}, the query processing times drop exponentially and that happens because the number of buckets is increased exponentially with the hash size value. \\
The evaluated accuracy in Fig. \ref{fig:experiment3} follows a linear growth, with the growth of the hash size for reasons describe above. In the plot, the growth isn't exactly uniform and that can be justified with two reasons: \begin{itemize}
    \item the results aren't very deterministic - although the margin of error of reported values isn't large, the values fluctuate because of the  random nature of the model.
    \item the improvement of accuracy throttles in high hash size values - the data can be sparse and unrelated to each other and there's only so much we can do to improve the model.
\end{itemize}
Euclidean distance has been used as a metric. It has shown to be able of proper accounting of distances. Alternatively, the hash values can be kept in bits of 1 and 0 and then the Hamming distance metric can be used to determine which buckets are closest to the current one. \\
Fig. \ref{fig:experiment4} is another accuracy plot where the distance relations for each of our testing samples are displayed. There we can see that the sequences are mainly monotonically increased.\\\\
The model has some limitations.
Our paper takes an agnostic approach to space complexity.
To improve querying times, a more efficient way to handle the retrieved candidates post-bucket collection could've been implemented. Different data structures could've been used. An analysis from an I/O complexity standpoint could be useful for this case. \\
Another issue that isn't addressed is that the width of the bucket which isn't fixed. The side effect of non fixed bucket widths is that sometimes random projections being really random might encapsulate the data in massively unbalanced buckets. In our experiments, we have found this could be the case if we are testing with a low number of samples (~100). On the other hand, a large number of samples (100k as in our case) ensures that the data is balanced enough that the average size of $2^{s}$ is achieved and the buckets are balanced, and since the true purpose of LSH models is high dimensional data, this issue can be disregarded. \\
Finally, the testing phase was impaired by a lack of 'ground truths' similar items that could've been used to directly compare the retrieved candidates.
 \section{Conclusions}
In this paper, we presented to you an approach for constructing a model which is able to query similar items in sub linear time. That is a tremendous improvement from a brute-force approach, which we could regard as impossible to be used in the experiment of this paper, as the dimensionality of the data is way too high. This implementation was done using pre-computed audio features. It can very easily be tweaked and use other features that describe the audio such as information about genre, the nature of the song and the singer. 
Our implementation lacks an opinion on the space complexity that an approach like this could have. Although, from the experiments done, this approach should be better even in that regard compared to other methods of near neighbor search such as the brute-force or some sort of dimensional trees. In the present times when we are dealing with big data, it’s very important to consider approaches like this that have already become the norm in solving similar items searching tasks.

\end{document}